\documentclass[10pt,journal,twoside]{IEEEtran}
\usepackage{cite}
\usepackage{amsmath,amssymb,amsthm,amsfonts}
\usepackage{mathrsfs}
\usepackage{mathtools}
\usepackage{nicefrac}
\usepackage{scalerel} 
\usepackage{verbatim}
\usepackage{algorithmic}
\usepackage{graphicx}
\usepackage{caption}
\usepackage{subcaption}

\usepackage{tabularx} 
\usepackage{booktabs}
\usepackage{array}
\usepackage{multirow}
\usepackage{csquotes}
\usepackage{pifont}
\usepackage{makecell}
\usepackage{textcomp}
\usepackage[usenames,dvipsnames]{xcolor}
\usepackage{underscore}
\usepackage{float}
\usepackage{threeparttable}
\usepackage{etoolbox,xspace}
\usepackage{adjustbox}
\usepackage{makecell}
\usepackage{tikz}
\usepackage{pgfplots}
\usepackage{pgfplotstable}
\usetikzlibrary{patterns}
\pgfplotsset{width=1\linewidth,compat=1.7}
\usetikzlibrary{calc,arrows,shapes,snakes,automata,backgrounds,petri,positioning}
\usepackage{orcidlink}
\usepackage[inline,shortlabels]{enumitem}
\usepackage{url}

\usepackage{breakurl}
\usepackage{hyperref}
\hypersetup{
    colorlinks=true,
    linkcolor=RedViolet,
    anchorcolor=black,
    citecolor=blue,
    urlcolor=black,
    breaklinks=true
}

\usepackage{letltxmacro}
\usepackage{xparse}

\makeatletter
\LetLtxMacro\ieeetran@appendix\appendix
\AtBeginDocument{%
	\RenewDocumentCommand{\appendix}{o}{%
		\IfValueTF{#1}{%
			\ieeetran@appendix[#1]%
		}{%
			\ieeetran@appendix%
		}%
	}
}
\makeatother

\usepackage[capitalise]{cleveref}
\crefname{section}{Sect.}{Sects.}
\crefname{equation}{}{}
\crefname{figure}{Fig.}{Figs.}
\Crefname{appendix}{Appx.}{Appx.}
\crefname{enumi}{}{}
\crefname{property}{Property}{Properties}
\crefname{lemma}{Lemma}{Lemmas}
\crefname{llemma}{Lemma}{Lemmas}
\creflabelformat{llemma}{#2\thectprop.#1#3}

\usepackage{stfloats}
\usepackage{CJKutf8}
\usepackage[utf8]{inputenc}

\usepackage[normalem]{ulem}  

\usepackage{pifont}

\def\BibTeX{{\rm B\kern-.05em{\sc i\kern-.025em b}\kern-.08em
    T\kern-.1667em\lower.7ex\hbox{E}\kern-.125emX}}

\makeatletter
\long\def\@IEEEtitleabstractindextextbox#1{\parbox{0.922\textwidth}{#1}}
\makeatother

\makeatletter
\def\@IEEEsectpunct{.\ \,}
\def\paragraph{\@startsection{paragraph}{4}{\z@}{1.5ex plus 1.5ex minus 0.5ex}%
	{0ex}{\normalfont\normalsize\sffamily\bfseries}}
\makeatother

\input{macros}

\begin{document}
\title{{\systemBold}: A Formally Verified Authentication Protocol for Multiprocessor Secure Boot under Hardware Supply-chain Attacks}
\author{
	Zhuoruo Zhang$^{\orcidlink{0000-0001-7896-1694}}$,
    Rui Chang$^{\orcidlink{0000-0002-0178-0171}}$,
	Mingshuai Chen$^{\orcidlink{0000-0001-9663-7441}}$,
    Wenbo Shen$^{\orcidlink{0000-0003-2899-6121}}$,\\
    Chenyang Yu$^{\orcidlink{0000-0002-7287-6971}}$,
	He Huang$^{\orcidlink{0009-0001-7718-7753}}$,
	Qinming Dai$^{\orcidlink{0009-0000-5549-0826}}$, 
    and
	Yongwang Zhao$^{\orcidlink{0000-0002-2284-1383}}$
 \thanks{The authors are with the Zhejiang University, Hangzhou 310027, China. Email: \{zhangzhuoruo, crix1021, m.chen, shenwenbo, 22151279, huanghe3776707, qinm\_dai,  zhaoyw\}@zju.edu.cn.}
}

\markboth{IEEE TRANSACTIONS ON INFORMATION FORENSICS AND SECURITY,~VOL.~XX, NO.~X}{PA-Boot: A Formally Verified Authentication Protocol for Multiprocessor Secure Boot under Hardware Supply-chain Attacks}


\maketitle

	\begin{abstract}
        Hardware supply-chain attacks are raising significant security threats to the boot process of multiprocessor systems. In this paper, we investigate critical stages of the multiprocessor system boot process and identify a new, prevalent hardware supply-chain attack surface that can bypass secure boot due to the absence of processor-authentication mechanisms. To defend against such attacks, in this paper, we present PA-Boot, the first formally verified processor-authentication protocol for secure boot in multiprocessor systems. PA-Boot is proved functionally correct and is guaranteed to detect multiple adversarial behaviors, such as processor replacements and man-in-the-middle attacks. The fine-grained formalization of PA-Boot and its fully mechanized security proofs are carried out in the Isabelle/HOL theorem prover with 306 lemmas/theorems and $\sim$7,100 LoC. We further implement in C an instance of PA-Boot. Experiments on the proof-of-concept implementation indicate that PA-Boot can effectively identify boot-process attacks with a considerably minor overhead (4.98\% on Linux boot process) and thereby improve the security of multiprocessor systems.
	\end{abstract}


\IEEEpeerreviewmaketitle

\setlength{\floatsep}{0.2\baselineskip}
\setlength{\textfloatsep}{0.2\baselineskip}
\setlength{\intextsep}{0.2\baselineskip}

\section{Introduction}\label{sec:introduction}
\IEEEPARstart{A}TTACKS during the boot process are notoriously hard to detect
because at this early stage of a device’s lifecycle, traditional countermeasures like firewalls and anti-viruses are not yet in place~\cite{tao2021dice}. A widely adopted defence against boot attacks is known as \emph{secure boot}~\cite{tygar1991dyad,DBLP:conf/isqed/Haj-YahyaWPBC19}, which enforces every boot stage to authenticate the subsequent stage such that only the firmware signed by an authorized entity (i.e., the device manufacturer) can be loaded and thereby establishes a \emph{chain-of-trust} in the entire boot process. During this process, the processors of a device serve as the \emph{root-of-trust} (RoT) to bootstrap the trust chain~\cite{DBLP:conf/isqed/Haj-YahyaWPBC19}. The authenticity of processors is thus of vital importance to system security.

However, the globalized and increasingly complicated hardware supply chains are threatening the trustworthiness of processors  -- the RoT in secure boot -- therefore exposing various modern devices to inevitable \emph{supply-chain attacks}~\cite{shwartz2017shattered,bohling2020subverting,skudlarek2016platform, christensen2020decaf,dhanuskodi2020counterfoil,han2021does,miller2013supply,nistsupply}. 
Specifically, many original equipment manufacturers (OEMs) nowadays outsource their hardware and/or firmware development to third-party suppliers without full inspection into their cybersecurity hygiene~\cite{shiralkar2021assessment,OEM1}, where the devices can be intercepted and implanted with compromised components during multiple hands of trade.
Such supply-chain attacks raise significant security threats and thereby an urgent request in identifying device vulnerabilities~\cite{spy2}, particularly, in the boot process~\cite{DBLP:journals/cacm/Frazelle20a,meadows2020chip}.

We focus on new hardware supply-chain attacks that can bypass secure boot of \emph{multiprocessor systems}.
A multiprocessor system includes a \emph{bootstrap processor} (BSP) responsible for initializing and booting the operating system and multiple \emph{application processors} (APs) activated after the operating system is up\footnote{The same convention applies to \emph{symmetric} multiprocessing (SMP): Whereas all the processors in an SMP system are considered functionally identical, they are distinguished as two types in the boot process.}.
As instances of hardware supply-chain attacks, an attacker can intercept a customer's multiprocessor device and either
\begin{enumerate*}[label=(\roman*)]
\item replace an AP with a compromised one, e.g., an AP with a pre-installed bootkit; or
\item implant an extra chip sabotaging the inter-processor communications (\emph{man-in-the-middle attacks}~\cite{DBLP:conf/sbcci/SantAnaMFM19}).
\end{enumerate*}
Such supply-chain attacks can give attackers control early in the boot process, allowing them to load a malicious bootloader or OS kernel. This enables high-privilege arbitrary code execution and evades detection by conventional security measures that rely on the OS.
For instance, see the proof-of-concept in Bloomberg's \enquote{Big Hack}~\cite{DBLP:journals/jetc/MehtaLPSRICWTA20,Trammell2018lecture}.

Existing research efforts focus on firmware integrity and provide no countermeasures against hardware supply-chain attacks in multiprocessor secure boot. Specifically, both the authenticity of APs and the inter-processor communications are conventionally \emph{trusted by default} universally across all modern multiprocessor systems. In fact, defending against this new hardware attack surface is challenging: It is difficult to examine all steps through the global supply chain from manufacturers to customers; moreover, identifying malicious components via hardware tampering detection techniques, e.g., circuit-based sensors and X-ray imaging, requires expertise and is time consuming~\cite{mosavirik2022impedanceverif, us-doc}.
Some work uses runtime monitors to record external behaviors of CPU chips (i.e., I/O, and memory read/write) and verifies chip integrity  \cite{zhu2021jintide}. However, specialized hardware components are required to extend the system.
It is thus desirable to equip the existing multiprocessor secure boot process with a mechanism for authenticating APs and securing communications without requiring custom hardware changes. 

In this paper, we present a processor-authentication protocol, called {\system}, to assure both the \emph{authenticity of APs} and the \emph{confidentiality of inter-processor communications} in the early stage of secure boot process for multiprocessor systems.
{\system} is capable of detecting multiple adversarial behaviors including AP replacements and man-in-the-middle attacks.
The boot process is aborted if any of the adversarial behavior is detected to prevent the attacker from taking control of the system.
The security and functional correctness of {\system} is verified based on deductive reasoning techniques: The formalization of {\system} and its fully mechanized security proofs (in terms of the AP authenticity, certificate integrity, etc.) are conducted in the (interactive) theorem prover Isabelle/HOL~\cite{nipkow2002isabelle}. 
This fine-grained formalization of {\system} in Isabelle/HOL succinctly captures its key components, the system behaviors, and a full range of adversarial capabilities against the protocol. To the best of our knowledge, \emph{{\system} is the first formally verified processor-authentication protocol for secure boot in multiprocessor systems}. We further implement in C an instance of {\system} called {\csystem}. Experiments simulated via ARM Fixed Virtual Platforms (FVP) suggest that {\csystem} can effectively identify multiple boot-process attacks -- by either manipulating the APs or tricking the AP-authentication mechanism -- with a considerably minor overhead and thus essentially improve the security of multiprocessor systems.
We have open-sourced our code, available at https://zenodo.org/records/14513558.

\paragraph*{\bf Contributions}
The main contributions are summarized as:
\begin{itemize}[leftmargin=16pt,labelwidth=8pt,labelsep=6pt]
    \item \textbf{Design of {\system}}: 
    We inspect critical steps in multiprocessor secure boot and identify a new, prevalent attack surface -- exhibiting hardware supply-chain attacks -- that may bypass secure boot due to the lack of AP authentication. To defend against such attacks, we design the first processor-authentication protocol {\system} for secure boot in multiprocessor systems that is amenable to formal verification via theorem proving.
    \item \textbf{Verification of {\system}}:
    We formalize {\system} based on multi-level abstraction-refinement in Isabelle/HOL via 91 locale/definitions.
    Meanwhile, we formalize multiple properties on both functional correctness and security (e.g.,  authenticity and integrity). 
    The proof that {\system} satisfies these properties is then fully mechanized in the form of 306 lemmas/theorems and $\sim$7,100 LoC
    \item \textbf{Implementation and evaluation}: 
    We implement {\csystem} in compliance with the formalization of {\system}.
    To build confidence in their consistency, we introduce a validation framework that extracts executable code from our Isabelle/HOL model and validates it against {\csystem}.
    We integrate {\csystem} in a real-world bootloader and show on FVP that {\csystem} can effectively identify multiple adversarial behaviors like replacing APs and man-in-the-middle attacks, with a considerably minor overhead (4.98\% on Linux boot).
\end{itemize}


\begin{figure}[t]
	\centering
	\includegraphics[width=\linewidth]{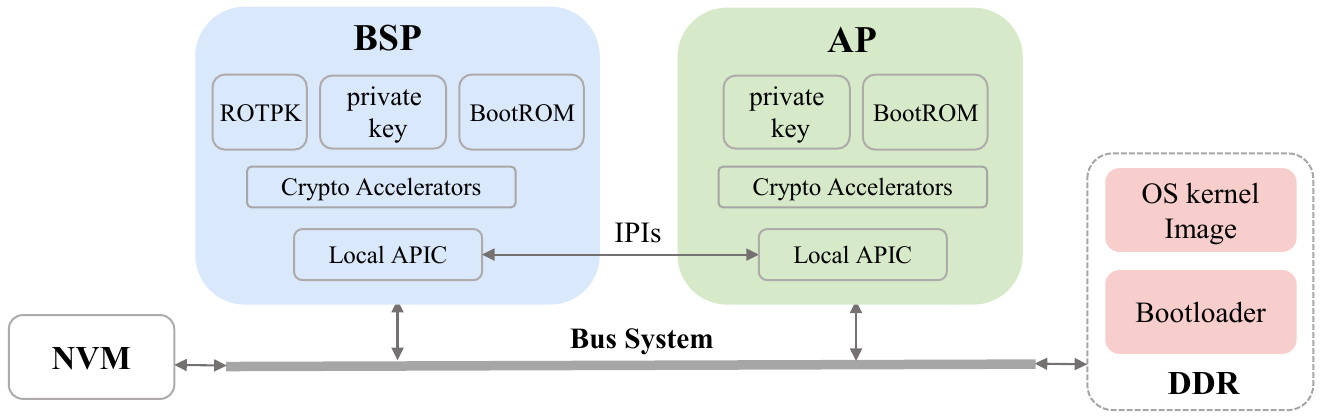}
	\caption{Key components in multiprocessor secure boot. }
	\label{fig:mp-arch1}
\end{figure}

\section{Background}\label{sec:background}

This section recapitulates secure boot in multiprocessor systems and hardware supply-chain attacks.

\subsection{Normal flow and limitation of secure boot}

Secure boot~\cite{tygar1991dyad} establishes a chain-of-trust (CoT) to ensure firmware integrity. It serves as a default (or often mandatory) feature in modern secure devices like laptops, desktops, smartphones, and IoT devices \cite{DBLP:conf/isqed/Haj-YahyaWPBC19, DBLP:conf/host/KhalidRI13}. 
The process begins with the immutable bootloader in read-only storage, considered the hardware root-of-trust (RoT). As the system's security foundation, the RoT must be immutable and tamper-resistant.
It initializes the hardware, locates the next boot image in Non-Volatile Memory (NVM), and loads it into memory. The RoT then verifies the image's integrity and authenticity using the Root of Trust Public Key (RoTPK), an OEM key stored immutably as a hash or in full.
The image, signed with the Root of Trust Private Key, ensures it comes from a trusted source. Execution proceeds only if the image is verified. Each subsequent boot stage follows this process, verifying the next image before transferring control. This chain-of-trust ensures that only verified components progress through the boot sequence, ultimately loading the kernel image securely.
Secure boot implementation varies across platforms \cite{marchand2023firmware}, with room for platform-specific operations. For enhanced security and efficiency, many systems integrate cryptographic engines for operations like signature verification and hash calculations.

Nevertheless, the secure boot's chain-of-trust (CoT) can be undermined by a malicious RoT. The integrity of the system depends on the assumption that the code in the first boot stage (acting as the RoT) is trustworthy or at worst buggy but non-malicious. Early systems stored this code on a write-protected flash memory (e.g., BIOS in PCs), but BIOS ROM is poorly protected and easily writable, leading to attacks exploiting the lack of access control during BIOS reflashing \cite{bios-bootkit}. In response, modern systems use hardwired bootROM, an immutable hardware component integrated into the CPU \cite{arm-boot, microsoft-riot}. Most research assumes the CPU chip, as part of the trusted computing base (TCB), is tamper-resistant and trustworthy, while off-chip components are vulnerable \cite{arm-boot, microsoft-riot, TCG}. However, we show that the hardware supply chain poses a risk, as malicious actors with access to the hardware supply chain can replace genuine CPU chips, compromising the RoT and gaining control over the system.

\subsection{Secure Boot in Multiprocessor Systems}\label{subsec:secure-boot}


\Cref{fig:mp-arch1} illustrates the key components of a multiprocessor secure boot system. 
Without loss of generality, we assume throughout the paper that the motherboard of a multiprocessor device is equipped with two processors, each in physically separate chip sockets: one as the BSP and the other as the AP. 
Each processor has a unique on-chip private key as its hardware identifier. The processors are connected via a shared bus to a mutable NVM and shared memory. Each processor includes a cryptographic extension module to accelerate operations. 
Except the shared bus, the BSP and AP communicate via a dedicated channel for inter-processor interrupts (IPIs). The motherboard is soldered a baseboard management controller (BMC), which is connected to the BSP and handles initialization and monitoring tasks. Note that the roles of BSP and AP are determined by their socket positions on motherboard, allowing the BMC to identify them accordingly.

The multiprocessor secure boot begins by executing the immutable code in the processors' bootROM, considered the hardware root-of-trust (RoT), which initializes the hardware. The AP enters a suspended state while the BSP proceeds with the following steps:
\begin{enumerate*}[label=(\roman*)] 
    \item Locating the bootloader image in the NVM, 
    \item Loading it into shared memory, 
    \item Verifying its authenticity and integrity using cryptographic accelerators, detecting any tampering with the certified images in NVM, and 
    \item Executing the image to load the next firmware stage. 
\end{enumerate*}
This authentication process propagates through all subsequent layers until the OS kernel is loaded. The BSP then broadcasts an IPI with the address for the next execution via its built-in \emph{local advanced programmable interrupt controllers} (LAPIC). The AP, which monitors the IPI channel, is activated and runs the OS kernel with the same privileges as the BSP. 
Variants of the typical multiprocessor architecture in \cref{fig:mp-arch1} may include, e.g., different inter-processor communication methods.

However, in the context of secure boot, as identified in our threat model below, an attacker with access to the hardware supply chain may bypass secure boot by exploiting either a compromised AP or an additional chip sabotaging the IPI channel and thus taking control of the target system.

\begin{figure*}[t]
	\centering
	\begin{tikzpicture}
		\draw (0, 0) node[inner sep=0] {\includegraphics[width = 0.8\linewidth]{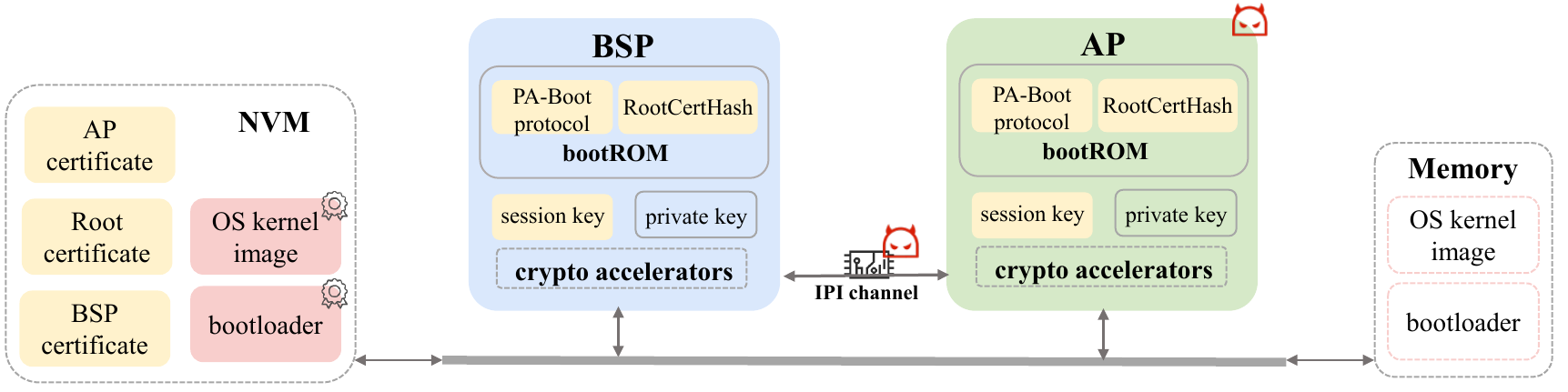}};
		\draw (4.61, 1.76) node {\tiny \cref{attack-AP}};
		\draw (1.35, -0.41) node {\tiny \cref{attack-chip}};
	\end{tikzpicture}
	\caption{Threat model of {\system}. {\devil} marks potential vulnerabilities to adversarial behaviors \cref{attack-AP} and \cref{attack-chip} in multiprocessor secure boot. Components of {\system} are colored in yellow. }
    \vspace{-1.5em}
	\label{fig:mp-arch2}
\end{figure*}

\section{Overview of Our Approach}\label{sec:overview}

This section identifies our threat model exhibiting the new attack surface in multiprocessor secure boot as hardware supply-chain attacks, and outlines our defense approach.

\subsection{Threat Model}
\label{subsec:threat}
In our threat model, we trust the CPU chip manufacturer (e.g., Qualcomm), PCB board manufacturer, and device OEM (e.g., Apple). 
In other words, the design and fabrication process of original CPU chips and boards are trusted, since they come from reliable hardware supply chains controlled by the OEM.
The OEMs can also take measures for tampering detection, like checking credentials, to ensure they receive genuine components from the vendors. 
In addition, attacks at the manufacturing steps can’t easily target a specific end-product, making attacks highly unlikely. 

We assume a realistic and powerful attacker who has physical access to the target multiprocessor device in the post-manufacturing hardware supply chain, i.e., in-transit from the OEM to the end-user or in the field. Therefore, the attacker can intercept the device and then replace the pluggable CPU chips on board and have malicious modification to the PCB, such as implanting a interposer chip on the board wires, to launch the following two \emph{attack vectors} (see {\devil} in \cref{fig:mp-arch2}) under no awareness of the BSP:

\begin{enumerate}[label=(\alph*), leftmargin=1.4em]
	\item\label{attack-AP} \emph{AP-replacement attack}: The attacker replaces the original AP with a malicious one that has, e.g., a factory-installed bootkit in its bootROM. Such a malicious AP can obtain via the shared memory bus secret data or high-privilege resources. It can also manipulate memory content and overwrite (parts of) the bootloader to be executed to take control of the system already at the boot stage.
	\item\label{attack-chip} \emph{Man-in-the-middle attack}: The attacker implants an extra interposer chip snooping on the IPI channel for inter-processor communications. The chip can sabotage traffic along the IPI channel by, e.g., substituting the memory entry pointer encoded in the activation IPI from BSP to AP, leading to control-flow hijacking and arbitrary code executions; moreover, it can sniff the IPI channel to capture secret data or interfere with runtime inter-processor communications concerning, e.g., remote TLB shootdowns.
\end{enumerate}

\paragraph*{\textbf{Assumptions}}
We make the following assumptions:
\begin{enumerate*}[label=(\roman*)]
\item The BSP is trusted. Its authenticity is ensured by the onboard baseboard management controller (BMC) before our protocol executes.
\item The on-chip bootROM is designed to be immutable \cite{zhao2014providing,arm-boot} and cannot be modified by an attacker, as modifying it after manufacture is exceptionally difficult \cite{zonenberg2025extraction} (see \cref{sec:discussion} for potential extension of our approach against very powerful attackers able to tampering with the bootROM). 
While the bootROM inside the AP remains immutable after manufacturing, as discussed in \cref{attack-AP}, attackers can more feasibly create a malicious AP by pre-installing malware during chip production. Later, in the post-manufacturing supply chain, they can replace the genuine AP plugged on device with the malicious one.
\item Even if an attacker introduces a malicious AP, they cannot extract or replicate the genuine AP’s on-chip private key. 
While advanced physical attacks using techniques like scanning electron microscopy (SEM) could theoretically target on-chip data, they are impractical due to their high demand for knowledge, time and equipment \cite{courbon2020practical}. 
Modern hardware protections mitigate these risks by binding the key to hardware identities \cite{tao2021dice} or intrinsic physical properties rather than storing it statically.
For example, techniques like 
Physically Unclonable Functions (PUFs) \cite{gassend2002silicon} regenerate keys at runtime using unique physical characteristics.

\item Cryptographic accelerators, like those used for elliptic-curve cryptography (ECC) and SHA-256 integrity checks, are trusted.
\end{enumerate*}

Note that hardware replacement/modification is a strong attack model, the adversary can physically tamper with all components on the device. Uniquely in our case, we only focus on the two new novel attack vectors. For instance, we do not consider physical attacks such as cold boot attacks or snooping attacks of the shared memory bus, given that such attacks have been studied and mitigated as in prior studies \cite{DBLP:conf/asplos/YoungNQ15,DBLP:conf/asplos/AwadMHSH16}.
Physical attacks other than internal hardware replacement, such as side-channel attacks are out of our scope.


\begin{figure*}[t]
	\centering
	\resizebox{.7\textwidth}{!}{
		\begin{tikzpicture}[
			block/.style args = {#1/#2/#3}{draw=#1, thick, align=flush center, rectangle, rounded corners, text width=#2, minimum height=#3},
			block/.default = gray/4.2cm/6mm
			]
			\tikzstyle{line} = [draw, thick, -latex']
			
			\begin{scope}
				\node[block=Maroon!60/4.4cm/6mm,fill=Maroon!10] (threat) {\phantom{\footnotesize{\S~\ref{subsec:threat}}}\hfill Threat model\hfill\footnotesize{\S~\ref{subsec:threat}}};
				
				\node[block=NavyBlue/4.4cm/39mm,fill=figblue!80,below=.6cm of threat] (pa-boot) {};
				\node[text width=4.4cm,above =-.6cm of pa-boot] (protocol)  {$\,${\system}\hfill\footnotesize{\S~\ref{sec:protocol}}};
				\node[block=gray/4cm/6mm,fill=figyellow!80,below=1.4cm of threat] (phase1) {\makecell{initiation phase\\[.08cm][certificates validation]}};
				\node[block=gray/4cm/6mm,fill=figyellow!80,below=.6cm of phase1] (phase2) {\makecell{\hspace*{-.2em}challenge-response phase\\[.08cm]\hspace*{-.2em}[AP authn.\ \& channel encr.]}};
				
				\node[block=ForestGreen/10.3cm/39mm,fill=figgreen!80,right=2.4cm of pa-boot] (isabelle) {};
				\node[text width=9.3cm,above =-.6cm of isabelle] (formalization)  {$\,$Formalization and verification in Isabelle/HOL\hfill~};
				\node[block=gray/4cm/6mm,fill=figyellow!80,right=2.8cm of phase1] (high-spec) {\phantom{\makecell{\footnotesize{\S~\ref{subsec:high-spec}}\\[.08cm]~}}\hfill\makecell{high-level\\[.08cm]specification $\spec_h$}\hfill\makecell{\footnotesize{\S~\ref{subsec:high-spec}\!}\\[.08cm]~}};
				\node[block=gray/4cm/6mm,fill=figyellow!80,right=1.67cm of high-spec] (low-spec) {\phantom{\makecell{\footnotesize{\S~\ref{subsec:low-spec}}\\[.08cm]~}}\hfill\makecell{low-level\\[.08cm]specification $\spec_l$}\hfill\makecell{\footnotesize{\S~\ref{subsec:low-spec}\!}\\[.08cm]~}};
				
				\node[block=gray/4cm/6mm,fill=figyellow!80,below=.6cm of high-spec] (high-prop) {\phantom{\;\:\makecell{\footnotesize{\S~\ref{subsec:hl-prop}}\\[.08cm]~}}\hfill\makecell{high-level\\[.08cm]properties}\hfill\makecell{\;\!\footnotesize{\S~\ref{subsec:hl-prop}\!\!\!\!\!\:}\\[.08cm]~}};
				\node[block=gray/4cm/6mm,fill=figyellow!80,below=.6cm of low-spec] (low-prop) {\phantom{\;\:\makecell{\footnotesize{\S~\ref{subsec:low-prop}}\\[.08cm]~}}\hfill\makecell{low-level\\[.08cm]properties}\hfill\makecell{\;\!\footnotesize{\S~\ref{subsec:low-prop}\!\!\!\!\!\;}\\[.08cm]~~\:\:\cmark\!\!\!}};
				
				
				
				\node[block=Orange!60/16.35cm/6mm,fill=Orange!10,below right=1cm and -4.65cm of pa-boot] (cpa-boot) {\phantom{\footnotesize{\S~\ref{sec:evaluation}}}\hfill Evaluation of {\csystem} (implementation in C) in terms of security and performance \hfill\footnotesize{\S~\ref{sec:evaluation}}};
				
				\coordinate[right=2.4cm of pa-boot] (isabelle-ghost);
				\coordinate[below=1cm of pa-boot] (cpaboot-ghost-left);
				\coordinate[below=1cm of isabelle] (cpaboot-ghost-right);
				
				\path [line] (threat) -- (pa-boot); 
				\path [line] (phase1) -- (phase2);
				
				\path [line] (high-spec) -- (low-spec) node [above,pos=0.5] {refined to};
				\path [line] (high-prop) -- (low-prop) node [above,pos=0.5] {refined to};
				\path [line] (high-spec) -- (high-prop) node [right,pos=0.5] {satisfies};
				\path [line] (low-spec) -- (low-prop) node [right,pos=0.5] {satisfies};
			
				\path [line] (pa-boot) -- (isabelle-ghost) node [above,pos=0.5] {protocol} node [below,pos=0.5] {adv.~behaviors};
				
				\path [line] (pa-boot) -- (cpaboot-ghost-left) node [right,pos=0.5] {instantiate};
				\path [line] (isabelle) -- (cpaboot-ghost-right) node [right,pos=0.5] {extract};
			\end{scope}
		\end{tikzpicture}
	}
	\caption{The workflow of our approach. {\cmark} indicates verified properties encoding security and functional correctness.}
	\label{fig:workflow}
        \vspace{-1.5em}
\end{figure*}

\subsection{Workflow of Our Approach}

\Cref{fig:workflow} sketches the overall workflow of our approach.
To defend against the hardware supply-chain attacks \cref{attack-AP,attack-chip} identified in our threat model, we develop the processor-authentication protocol {\system}, which orchestrates the secure boot process via an \emph{initiation phase} to validate the processor certificates and a \emph{challenge-response phase} to authenticate the AP identity and to secure the inter-processor communication channel. 
We then formalize it and the possible (adversarial and normal) behaviors thereof in Isabelle/HOL 
as a \emph{high-level specification} $\spec_h$ and a set of \emph{high-level properties} (encoding security and functional correctness). These high-level ingredients -- capturing the core components of {\system} -- are further \emph{refined} into their low-level counterparts, i.e., the \emph{low-level specification} $\spec_l$ and a set of \emph{low-level properties}. We then conduct a fully mechanized proof via theorem proving that $\spec_l$ satisfies the low-level properties. 
Finally, we derive an implementation of {\system} from the formalized model as {\csystem} based on a code-to-spec review~\cite{zhao2017refinement}. We show that {\csystem} effectively identifies various boot-process attacks 
with a considerably minor overhead and thereby improves the security of multiprocessor systems.

Separating a protocol’s formalization into high- and low-level specifications is a common strategy for abstraction and refinement (e.g., \cite{DBLP:conf/sosp/HawblitzelHKLPR15,zhao2017refinement}). The high-level specification $\spec_h$ provides a simple system behavior description, while the low-level specification $\spec_l$, closer to implementation, details all possible protocol executions and attacker interactions.
We verify security properties and functional correctness using $\spec_l$, as it captures specific protocol configurations and execution traces. 
$\spec_h$ helps to gain confidence in the correctness of the system’s specification and its alignment with expectations. 
Its abstract, declarative nature reduces specification errors and facilitates their discovery. The refinement proof between $\spec_l$ and $\spec_h$ further extends that confidence to $\spec_l$.

\begin{table}[t]\small
	\centering
	\renewcommand{\arraystretch}{1.2}
	\caption{Frequently used notations.}
	\label{tab:notations}
		\begin{tabularx}{\linewidth}{ccX}
			\toprule
			\textbf{Notation} &&\textbf{Intuitive meaning} \\
			\midrule
                $\certroot$/$\certbsp$/$\certap$ && certificate of root/BSP/AP, resp.\\
			$\hashrc$ && hash value of $\certroot$\\
			$\kpairbsp$ && public-private key pair of BSP\\
			$\kpairap$ && public-private key pair of AP\\
			$\noncecha$/$\nonceresp$ && nonce generated by AP/BSP, resp.\\
   			$\ekpairbsp$ && ephemeral public-private key pair generated by BSP for $\sessk$\\
			$\ekpairap$ && ephemeral public-private key pair generated by AP for $\sessk$\\
			$\sessk$ && shared session key for the secure channel eventually established between\ BSP and AP \\
			\bottomrule
		\end{tabularx}
\end{table}

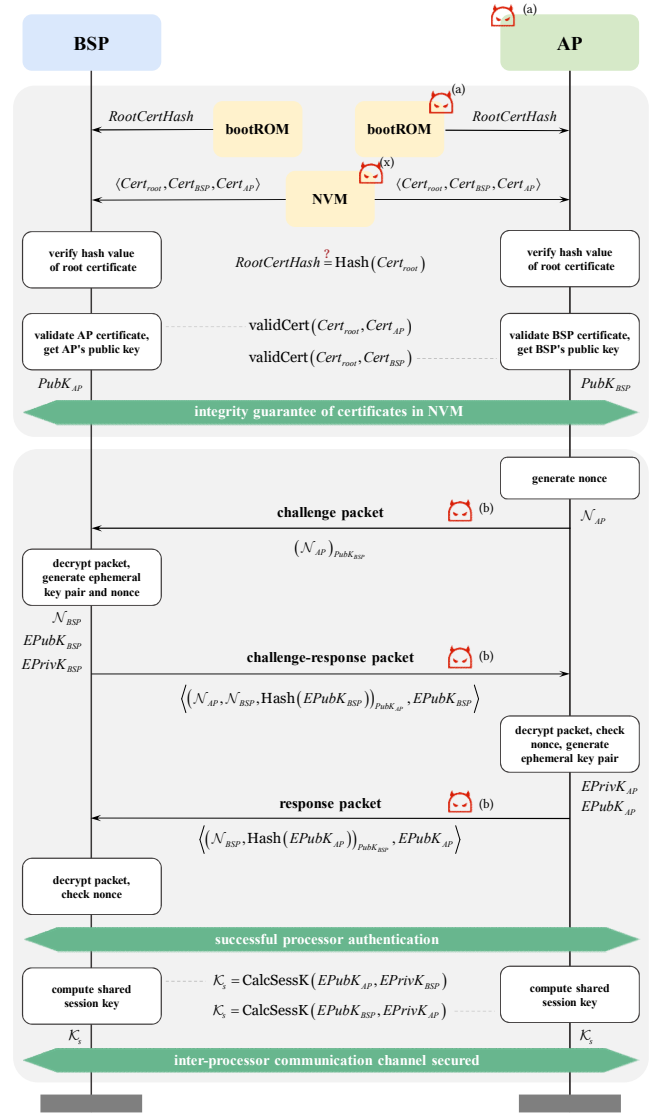
\begin{figure}[t]
	\centering
	\begin{tikzpicture}
		\draw (0, 0) node[inner sep=0] {\includegraphics[width = 1\linewidth]{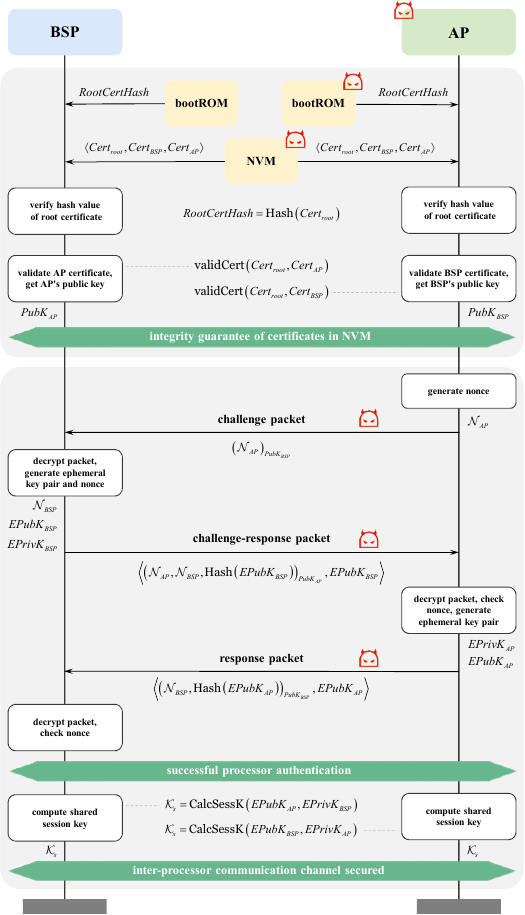}};
		\draw (-0.06, 4.06) node {\tiny $\qmark$};
		\draw (2.64, 7.60) node {\tiny \cref{attack-AP}};
		\draw (1.8, 6.4) node {\tiny \cref{attack-AP}};
		\draw (2.06, .66) node {\tiny \cref{attack-chip}};
		\draw (2.06, -1.31) node {\tiny \cref{attack-chip}};
		\draw (2.06, -3.28) node {\tiny \cref{attack-chip}};
	\end{tikzpicture}
	\caption{Key message flows in {\system}. {\devil} marks potential vulnerabilities to adversarial behaviors \cref{attack-AP} and \cref{attack-chip}.}
	\label{fig:message-flow}
\end{figure}

\section{Design of {\systemBold}}\label{sec:protocol}

This section details the design of our processor authentication protocol {\system}. It augments multiprocessor secure boot with several key components as depicted in \cref{fig:mp-arch2}. 
{\system} steers the secure boot process in a certificate-based, two-phase manner: The processor certificates are validated in the \emph{initiation phase} and thereafter, in the \emph{challenge-response phase}, the AP identity is authenticated and the inter-processor communication channel is encrypted. 
We first explain the necessary operations to set up the stage for running {\system} and then elucidate key message flows in the abovementioned two phases. 
Frequently used notations are collected in \cref{tab:notations}.

Our protocol requires each processor to request its certificate from the certificate authority (CA) at the OEMs and store the signed certificate once issued.
More concretely, the OEM first generates authorized certificates $\cert{BSP}$ and $\cert{AP}$ respectively by signing $\pubkbsp$ and $\pubkap$ with its own private key (these certificates thus can be validated by the OEM's public key in $\cert{root}$), and then stores all necessary certificates $\left\langle\cert{root}, \cert{BSP}, \cert{AP}\right\rangle$ in the NVM. 
Meanwhile, the hash value of the root certificate $\certroot$ is stored in the bootROM of the processors. Recall that any attacker that can compromise the BSP's bootROM is beyond the scope of this paper.


   \begin{remark}
	{\system} is \emph{not} tailored to dual-processor devices, rather, it applies to the setting of \emph{multiple} APs where the BSP can authenticate these APs in sequence.
	\qedT
\end{remark}

\subsection{The Initiation Phase}\label{subsec:initiation}

The two processors BSP and AP behave \emph{symmetrically} in the initiation phase: As depicted in (the upper part of) \cref{fig:message-flow}, each processor first 
reads the precomputed hash value $\hashrc$ of $\certroot$ from its bootROM as well as the chain of certificates $\left\langle\cert{root}, \cert{BSP}, \cert{AP}\right\rangle$ from the NVM.
It then checks the validity and integrity of $\cert{root}$, namely, checking whether $\hashrc = \hash{\cert{root}}$. If this is indeed the case, then $\cert{root}$ is used to validate the certificate of the other processor by applying the function $\verifycert{\cert{root}}{\cdot}$, 
which further reveals the public key of the other processor.
The non-volatile memory (NVM) is easily modifiable, giving the attacker the ability to alter its contents. The attacker could replace legitimate certificates with those associated with a malicious AP, thereby attempting to deceive {\system}. However, since the forged certificate is not signed by a trusted certificate authority (CA), the system can detect its invalidity during the certificate validation process.


\subsection{The Challenge-Response Phase} \label{subsec:chal-resp}

As depicted in (the lower part of) \cref{fig:message-flow}, once confirming the validity of $\cert{BSP}$ in the initiation phase, the AP sends -- via the inter-processor communication channel -- a \emph{challenge packet} $\left(\nonce{AP}\right)_{\pubk{BSP}}$, that is, a randomly generated nonce $\nonce{AP}$ encrypted with $\pubk{BSP}$. Note that a typical nonce is a 32-byte random number which is practically infeasible to guess. Upon receiving this packet, the BSP decrypts the packet via $\privk{BSP}$ 
to get $\nonce{AP}$ and then sends a \emph{challenge-response packet}
\begin{align*}
	\left\langle \left( \nonce{AP},\, \nonce{BSP},\, \hash{\epubk{BSP}} \right)_{\pubk{AP}},\: \epubk{BSP} \right\rangle
\end{align*}%
where $\nonce{AP}$ is the challenge nonce generated by AP, $\nonce{BSP}$ is the response nonce generated by BSP, and $\epubk{BSP}$ is the ephemeral public key generated by BSP (for computing the shared session key $\sessk$ later); this challenge-response packet is encrypted by $\pubk{AP}$ except for the $\epubk{BSP}$ part. After receiving the challenge-response packet, the AP first decrypts the packet and checks its integrity, i.e., checking whether the received $\nonce{AP}$ is identical to the previously generated one \emph{and} whether the decoded $\hash{\epubk{BSP}}$ is identical to the hash value of the received $\epubk{BSP}$. 
If this is indeed the case, the AP stores $\epubk{BSP}$ and generates its ephemeral key pair $\ekpairap$. Then, analogously, the AP responds to the BSP with a \emph{response packet}
\begin{align*}
	\left\langle \left( \nonce{BSP},\, \hash{\epubk{AP}} \right)_{\pubk{BSP}},\: \epubk{AP} \right\rangle~.
\end{align*}%
The BSP then decrypts the packet and checks its integrity by similar means as mentioned above.

These message flows by now in the challenge-response phase are able to detect the aforementioned attack vector \cref{attack-AP}, i.e., AP replacements, and \cref{attack-chip} man-in-the middle attacks attempting to sabotage the communication packets. To protect the communication channel from future interference, {\system} further produces a \emph{shared session key} $\sessk$ such that subsequent inter-processor communications can be secured using $\sessk$. This concludes the entire challenge-response phase of {\system}. We remark that the (symmetric) shared session key $\sessk$ can be calculated by BSP (resp.\ AP) based on $\epubk{AP}$ and $\eprivkbsp$ (resp.\ $\epubk{BSP}$ and $\eprivkap$) using the Diffie-Hellman (DH) key exchange algorithm~\cite{diffie1976new}. 

\section{Formalization in Isabelle/HOL}\label{sec:formalization}

This section presents the formalization of {\system} in the theorem prover Isabelle/HOL~\cite{nipkow2002isabelle}. The formal model consists of a \emph{high-level specification} $\spec_h$ and a refined \emph{low-level specification} $\spec_l$.
$\spec_h$ captures core components of {\system} and gives the simplest description of the system behavior, whereas $\spec_l$ is a \emph{symbolic model} closer to the implementation layer, encoding a more fine-grained characterization of all possible executions of the system.
We opt for deductive verification as implemented in Isabelle/HOL due to its scalability and inherent support of abstraction refinement~\cite{DBLP:conf/mkm/Ballarin06} and code generation~\cite{codegen}. 

\paragraph*{\textbf Adversary model}
For the adversarial behavior \cref{attack-AP} identified in \cref{subsec:threat}, we consider the possibility that the agent AP is compromised, allowing the compromised agent to manipulate its long-term keys.
For the adversarial behavior \cref{attack-chip} identified in \cref{subsec:threat}, we explicitly model a classical Dolev-Yao-style adversary \cite{DBLP:conf/focs/DolevY81} who
 has full control over the insecure BSP-AP communication channel.
However, the adversary is limited by the constraints of the cryptographic methods used: he cannot forge signatures or decrypt messages without knowing the key (the black box cryptography assumption).

\subsection{High-Level Specification}
\label{subsec:high-spec}

The high-level specification $\spec_h$ of {\system} is encoded as a finite-state \emph{acyclic labelled transition system} representing the protocol executions under certain security contexts:
\begin{definition}[High-Level Specification of {\systemBold}]\label{def:high-spec}
	The \emph{high-level specification of {\system}} is a quintuple
	\begin{align*}
		\spec_h \ddefeq \!\left\langle \states, \inistate, \idlstate, \labels, \trans \right\rangle~, \quad \text{where}
	\end{align*}%
	\begin{itemize}[leftmargin=16pt,labelwidth=8pt,labelsep=6pt]
		\item $\states$ is a finite set of \emph{states} encoding the protocol configurations,
		\item $\inistate \in \states$ is the \emph{initial state},
		\item $\idlstate \in \states$ is the \emph{ideal state} signifying attack-free authentication, 
		\item $\labels$ is a finite set of \emph{event labels} representing actions of the processors and the attacker, and
		\item $\trans \subseteq \states \times \labels \times \states$ is a finite set of \emph{labelled transitions}.
	\end{itemize}
\end{definition}

A labelled transition $\tran = (s, \alpha, s') \in \trans$, denoted by $\step{s}{\alpha}{s'}$, yields a jump from the \emph{source state} $s$ to the \emph{target state} $s'$ on the occurrence of event $\alpha$. We consider deterministic transitions, i.e., if $\step{s}{\alpha}{s'}, \step{s}{\alpha}{s''}$ are both transitions in $\trans$, then $s' = s''$. 
The set of \emph{terminal states} is defined as $\stable \defeq \{s \in \states \mid \forall \alpha \in \labels.\: \nexists s' \in \states \colon  s\neq s' \wedge (s, \alpha, s') \in \trans \}$, i.e, a state $s$ is \emph{terminal} iff $s$ has no successors other than itself. Note that the ideal state $\idlstate$ is necessarily a terminal state in $\stable$.
A \emph{run} $\run$ of $\spec_h$, denoted by $\steps{s_0}{A}{s_n}$ with $A = \alpha_1 \alpha_2 \cdots \alpha_n \in \labseq$, is a finite sequence $\run = \step{s_0}{\alpha_1}{\step{s_1}{\alpha_2}{\step{\cdots}{\alpha_n}{s_n}}}$ with $s_n \in \stable$, $\step{s_{i-1}}{\alpha_i}{s_i} \in \trans$ for any $1 \leq i \leq n$.
We denote by $\runs$ the set of all possible runs of $\spec_h$. Given a run $\run = \steps{s_0}{A}{s_n} \in \runs$, $\tail{\run}$ denotes the tail state $s_n$ of $\run$. 

\paragraph*{\textbf Security contexts and indicator functions}
Every state $s \in \states$ of $\spec_h$ encodes, amongst others, the current \emph{security context} $c \in \configs$ consisting of security-related system configurations of the underlying \emph{security assets}, including the two processors, the NVM, and the inter-processor communication channel. We apply the function $\context\colon \states \to \configs$ to extract the security context pertaining to a state; for simplicity, we write $\context_s$ as shorthand for $\context(s)$. Let $\bools \defeq \{\TRUE, \FALSE\}$. We employ two indicator functions to witness the presence/absence of security threats: The (partial) \emph{benignity function} $\benign\colon \configs \rightharpoonup \bools$ signifies whether a security context $c$ is benign (in the state that $c$ is associated with), i.e., whether the AP and the certificates in the NVM are genuine; the \emph{free-of-attack function} $\mitm\colon \runs \to \bools$ determines whether a man-in-the-middle attack has \emph{not} occurred along a run of $\spec_h$, i.e., during one possible execution of the protocol. 

$\spec_h$ together with, e.g., its indicator functions are formalized as a \emph{locale module}~\cite{DBLP:conf/mkm/Ballarin06} in Isabelle/HOL -- an emerging mechanism for abstract reasoning -- consisting of abstract types and primitive operations that can be interpreted in different contexts. 
We omit the detailed locale formulation here.

\begin{center}
	\begin{table}[t]
		\renewcommand\arraystretch{1.3}
		\caption{The set $\labels$ of events in $\spec_l$.}
		\centering
		\label{tab:behaviors}
		\begin{tabularx}{\linewidth}{cll}
			\toprule
			\textbf{ID} & \textbf{Name} & \textbf{Description of the event} \\ 
			\midrule
			\multicolumn{3}{l}{\textbf{Processor behaviors (executed by BSP or AP)}}\\[.1cm]
			
			{\scriptsize\newcircled{\text{1}}}& Read\_ROM & read $\hashrc$ in bootROM\\  
			
			{\scriptsize\newcircled{\text{2}}}& Read\_NVM & read $\left\langle\cert{root}, \cert{BSP}, \cert{AP}\right\rangle$ in NVM\\
			
			{\scriptsize\newcircled{\text{3}}}& Verify\_RCHash & verify if $\hashrc = \hash{\cert{root}}$\\
			
			{\scriptsize\newcircled{\text{4}}}& Verify\_Cert & verify the other processor's cert.\ via $\cert{root}$\\
			
			{\scriptsize\newcircled{\text{5}}}& Gen\_Nonce & generate a fresh nonce\\
			
			{\scriptsize\newcircled{\text{6}}}& Send\_Packet & send a packet to the channel\\
			
			{\scriptsize\newcircled{\text{7}}}& Receive\_Packet & receive a packet via the channel\\
			
			{\scriptsize\newcircled{\text{8}}}& Parse\_Packet & decrypt a packet and check its integrity\\
			
			{\scriptsize\newcircled{\text{9}}}& Gen\_EpheKey & generate ephemeral public-private key pair\\
			
			{\scriptsize\circled{\text{10}}}& Gen\_SessKey & calculate $\mathcal{K}_{\textit{s}}$ to secure the comm.\ channel\\
			\midrule
			\multicolumn{3}{l}{\textbf{Adversarial behavior (executed by the attacker)}}\\[.1cm]
			{\scriptsize\circled{\text{11}}} & Attack & perform a man-in-the-middle attack\\
			\bottomrule
		\end{tabularx}
	\end{table}
\end{center}
\vspace{-2em}

\subsection{Low-Level Specification}
\label{subsec:low-spec}

Next, we refine $\spec_h$ of {\system} to its low-level counterpart $\spec_l$ by instantiating the state space as concrete protocol configurations and the labelled transitions as event-triggered actions of the processors or the attacker. The fact that \emph{$\spec_l$ refines $\spec_h$}, written as $\spec_h \sqsubseteq \spec_l$, is proved in Isabelle/HOL by interpreting locales as parametric theory modules~\cite{DBLP:conf/mkm/Ballarin06}.

\begin{figure*}
    \centering
    \begin{subfigure}{0.35\textwidth}
        \centering
        \includegraphics[width=\linewidth]{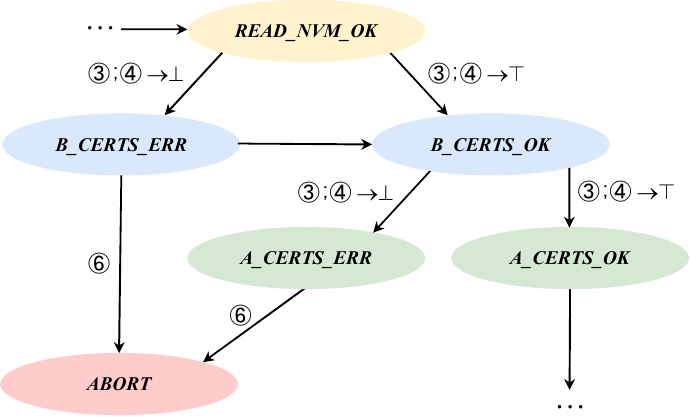}
        \caption{validating NVM certificates}
        \label{subfig:partstate-certs}
    \end{subfigure}%
    \hspace{0.1\linewidth}
    \begin{subfigure}{0.35\textwidth}
        \centering
        \includegraphics[width=\linewidth]{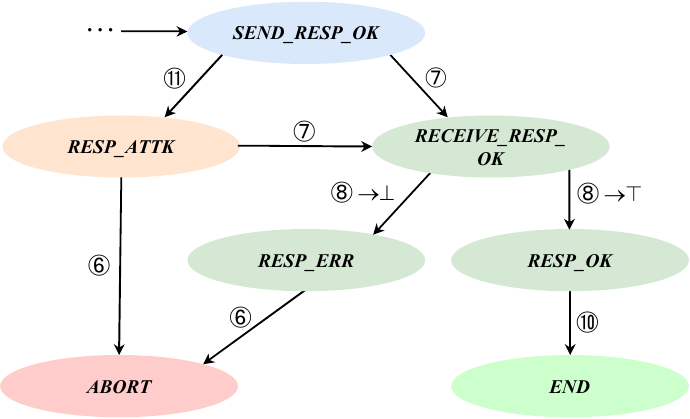}
        \caption{attack on the resp.\ packet}\label{subfig:partstate-resp}
    \end{subfigure}
    \caption{Snippets of the state machine $\spec_l$ of {\system}. A state $s$ is identified by its status and colored in light green (normal termination), red (abnormal termination), or in blue, dark green, yellow, or orange if the transition to $s$ is triggered by BSP, AP, both BSP and AP, or the attacker, respectively; The (condensed) transition {\footnotesize{$\eventp;\eventq; \ldots \!\rightarrow\! \top$}}, for $p,q \in \{3, 4, 8\}$ (cf.\ \cref{tab:behaviors}), is triggered if all the checks in the event sequence {\footnotesize{$\eventp,\eventq,\ldots$}} are successful, otherwise {\footnotesize{$\eventp;\eventq; \ldots \!\rightarrow\! \bot$}} is triggered.}
	\label{fig:partstate}
 \vspace{-1.5em}
\end{figure*}

A state in $\spec_l$ is encoded as a \emph{record} construct in Isabelle/HOL collecting all fields related to the protocol configuration:
\begin{align*}
	\record\,~\type{State} \eeq \{\ &\field{bsp} \dblcolon \type{Processor},~~\field{ap} \dblcolon \type{Processor},\\
	 &\field{env} \dblcolon \type{Envir},~~\field{status} \dblcolon \type{Status} \ \}~.
\end{align*}%
%
Specifically, $\type{Processor}$ encapsulates all the BSP/AP-related ingredients in {\system} (cf.\ \cref{fig:message-flow}); $\type{Envir}$ encodes the \emph{environment} of the processors, i.e., the NVM storing certificates
and the inter-processor channel carrying communication packets; 
$\type{Status}$ signifies the current status of {\system}, which can be $\type{INIT}$ (initialization), $\type{OK}$ (normal execution), $\type{ERR}$ (failure in certificate validation or packet parsing), 
$\type{ATTK}$ (presence of man-in-the middle attacks), $\type{END}$ (normal termination), and $\type{ABORT}$ (abnormal termination). 
In particular, a man-in-the middle attack will be recognized by {\system} when parsing the attacked communication packet and thus leads to an $\type{ERR}$ state; moreover, once a run visits an $\type{ERR}$ state, it raises an error-specific alarm and terminates in the (unique) $\type{ABORT}$ state. In fact, $\type{END}$ and $\type{ABORT}$ represent terminal states $\stable$ as defined in $\spec_h$, where $\type{END}$ particularly marks the ideal state $\idlstate$. 
Note that $\type{OK}$, $\type{ERR}$, and $\type{ATTK}$ are refined to more fine-grained status types using prefixing, see examples in \cref{fig:partstate}.


As listed in \cref{tab:behaviors}, the set $\labels$ of event labels in $\spec_h$ is instantiated in $\spec_l$ as concrete actions of the protocol participants. 
The set $\trans$ of event-triggered transitions in $\spec_l$ is specified based on \type{State} and $\labels$. \cref{fig:partstate} depicts parts of the state machine $\spec_l$, which represent typical scenarios in {\system}, e.g., validating NVM certificates (\cref{subfig:partstate-certs}) and detecting AP replacements or man-in-the-middle attacks on the response packet (\cref{subfig:partstate-resp}). 

The security-related utility functions $\context$, $\benign$ and $\mitm$ as described in the high-level specification are instantiated in the lower-level counterpart as well. For example, for a state $s$ in $\spec_l$, $\context_s = \context(s)$ yields the security context associated with $s$, that is, the components $\langle \field{bsp}, \field{ap}, \field{env} \rangle$ in $s$; moreover, given a run $\run$ of $\spec_l$, $\mitm(\run)$ determines whether or not a man-in-the-middle attack, i.e., {\scriptsize{\circled{\text{11}}}}, is absent along $\run$.

\section{Formal Verification}\label{sec:verification}

This section presents the mechanized proof in Isabelle/HOL that {\system} is functionally correct and suffices to detect aforementioned adversarial behaviors \cref{attack-AP} and \cref{attack-chip}. The core ingredient to the proof is the formalization of the corresponding properties on functional correctness and security -- first at the level of $\spec_h$ and then at $\spec_l$ in a refined form. 

\subsection{High-Level Properties}
\label{subsec:hl-prop}

\paragraph*{\textbf Functional correctness}
We require {\system} to be \emph{functionally correct}, in the sense that the benignity of the security context remains \emph{invariant} during any possible execution of {\system}:
\begin{property}[Functional Correctness w.r.t.\ $\spec_h$]\label{prop:invariant-high}
	\[\forall ~\run = \steps{s_0}{A}{s_n} \in \runs,~i\in[1,n].~ \benign\left(\context_{s_i}\right) \iff \benign\left(\context_{s_0}\right) \tag{$\dagger$}\]
\end{property}%
\noindent
Recall that $\benign(\context_{s_0}) = \TRUE$ iff the AP and the certificates in the NVM are genuine upon the start of {\system}, thereby witnessing the absence of adversarial behaviors \cref{attack-AP}. Provided a reasonable assumption that these two adversarial behaviors can only be conducted during the hardware supply chain, i.e., before the launch of {\system}, \cref{prop:invariant-high} ensures that our protocol per se does not alter the benignity of the security context during all of its possible executions.


\paragraph*{\textbf Security property}
{\system} should be able to \emph{secure the boot process}, in the sense that the successful authentication by {\system} (indicated by reaching the ideal state $\idlstate$) guarantees that every boot flow thereafter involves only trusted security assets and is free of man-in-the-middle attacks:
%
\begin{property}[Security w.r.t.\ $\spec_h$]\label{prop:security-high}
        \vspace{-0.5em}
        \[\forall ~\run \in \runs. ~ \tail{\run} = \idlstate \iff \benign\left(\context_{s_0}\right) \wedge \mitm\left(\run\right) \tag{$\ddagger$}\]
         \vspace{-2em}
\end{property}%
\noindent
Intuitively, \cref{prop:security-high} ensures that a run $\run$ of $\spec_h$ terminates in the ideal state $\idlstate \in \stable$ if and only if $\mitm(\pi) = \TRUE$ and $\benign(\context_{s_0}) = \TRUE$ (so is $\benign(\context_{s})$ for any $s$ along $\run$, due to \cref{prop:invariant-high}), that is, $\run$ is free of man-in-the-middle attacks and both the AP and the certificates in the NVM remain genuine along $\run$. Moreover, the established shared session key $\sessk$ ensures that {\system} is able to secure the entire boot process.

\subsection{Low-Level Properties} \label{subsec:low-prop}

Next, \cref{prop:invariant-high,prop:security-high} are refined w.r.t.\ $\spec_l$ in the form of Isabelle/HOL lemmas. We use $\mydot\!$ to extract specific fields of a state in $\spec_l$, e.g., $s\mydot\field{ap}\mydot\field{private\_key}$ denotes $\privkap$ in $s$.

\paragraph*{\textbf Functional correctness}
At the level of $\spec_l$, \cref{prop:invariant-high} boils down naturally to the requirement that the private keys of the two processors, $\hashrc$ stored in their bootROMs, and the certificates stored in the NVM remain unchanged during all possible executions of {\system}:
%
%
\begin{lemma}[Functional Correctness w.r.t.\ $\spec_l$]\label{lem:invariant-low}
	\begin{align*}
        \forall~\run &= \steps{s_0}{A}{s_n} \in \runs,~i\in[1,n].~  s_i\mydot\field{env}\mydot\field{nvm} = s_0\mydot\field{env}\mydot\field{nvm} \\\wwedge
		&s_i\mydot\field{ap}\mydot\field{private\_key} = s_0\mydot\field{ap}\mydot\field{private\_key} \\\wwedge
		&s_i\mydot\field{bsp}\mydot\field{private\_key} = s_0\mydot\field{bsp}\mydot\field{private\_key} \\\wwedge
		&s_i\mydot\field{ap}\mydot\field{root\_cert\_hash} = s_0\mydot\field{ap}\mydot\field{root\_cert\_hash} \\\wwedge
		&s_i\mydot\field{bsp}\mydot\field{root\_cert\_hash} = s_0\mydot\field{bsp}\mydot\field{root\_cert\_hash} \\[-1em]
    \end{align*}%
\end{lemma}


\paragraph*{\textbf Security properties}
In the presence of adversarial behaviors \cref{attack-AP} and \cref{attack-chip}, we aim to establish -- by the end of protocol execution -- three types of security goals of the underlying security assets at the level of $\spec_l$:
\begin{itemize}[leftmargin=16pt,labelwidth=8pt,labelsep=6pt]
    \item \emph{Authenticity} (against \cref{attack-AP}): The identity of AP is validated.
    \item \emph{Integrity} (against \cref{attack-AP} and \cref{attack-chip}): The inter-processor communication packets and the certificates in NVM remain unmodified (even before the execution of {\system}).
    \item \emph{Confidentiality} (against \cref{attack-chip} for future executions): The established shared session key $\sessk$ (for encrypting future communication messages) is known only to BSP and AP.
\end{itemize}
The above-mentioned security goals are encoded as three individual lemmas interpreting different protocol-execution scenarios. Specifically, the protocol under normal execution (cf.\ \cref{lem:security-low-norm}) should eventually terminate in the $\type{END}$ state where the processors are mutually authenticated and $\sessk$ is established; otherwise, if an adversarial behavior that violates either authenticity or certificate integrity is identified -- i.e., the AP has been replaced or the certificates in NVM have been tampered with (\cref{lem:security-low-B}), or an inter-processor communication packet has been modified by a man-in-the-middle attack (cf.\ \cref{lem:security-low-M}) -- then the protocol should raise error-specific alarms by visiting the $\type{ERR}$ states and eventually terminate in the $\type{ABORT}$ state.

\setcounter{ctprop}{2}
%
\begin{llemma}[Security w.r.t.\ $\spec_l$ under Normal Executions]\label{lem:security-low-norm}
\vspace{-0.5em}
\[
  \begin{split}
    \forall~\run \in \runs.~ 
    \benign\left(\context_{s_0}\right) \wedge \mitm\left(\run\right) \implies \tail{\run}\mydot\field{status} = \type{END}~\\[-0.4em]
\end{split}
\]
\end{llemma}%
\noindent
\cref{lem:security-low-norm} declares that starting from the initial state $s_0$ where the associated security context is benign, if there is no man-in-the-middle attack during the execution, {\system} must terminate in the $\type{END}$ state signifying attack-free authentication and the establishment of $\sessk$.



\begin{llemma}[Security w.r.t.\ $\spec_l$ against Tampered Configurations]\label{lem:security-low-B}
\vspace{-0.5em}
\[
  \begin{split}
 &\forall~\run = \steps{s_0}{A}{s_n} \in \runs.~\exists~ i\in[1,n].~\neg\benign\left(\context_{s_0}\right) \implies \\
        &\bigl(\,\tail{\run}\mydot\field{status} = \type{ABORT} \wwedge
       s_i\mydot\field{status} = \type{\%_ERR}\  \,\bigr)\\
       &\textnormal{where}~ \type{\%}~ \textnormal{matches}~ \type{A_CERTS},~\type{B_CERTS} ~\textnormal{or} ~\type{RESP}, ~\textnormal{see}~ \textnormal{\cref{fig:partstate}}\\[-0.4em]
\end{split}
\]
\end{llemma}%
%
\noindent
\cref{lem:security-low-B} states that in case $\context_{s_0}$ is compromised (due to adversarial behaviors \cref{attack-AP} and attacker's possible modification to necessary certificates in the NVM to impersonate a benign AP), {\system} should raise an error-specific alarm by visiting the corresponding $\type{ERR}$ state 
until it eventually terminates in the $\type{ABORT}$ state.


\begin{llemma}[Security w.r.t.\ $\spec_l$ against Man-in-the-Middle Attacks]\label{lem:security-low-M}
\vspace{-0.5em}
 \[
 \begin{split}
      &\qquad \forall~\run = \steps{s_0}{A}{s_n} \in \runs.~\exists~ i,j\in[1,n] ~s.t.~ i<j.\\
      &\qquad \neg\mitm\left(\run\right) \implies
       \bigl(\,\tail{\run}\mydot\field{status} = \type{ABORT} \wwedge \\
        &\qquad s_i\mydot\field{status} = \type{\%_ATTK} \wwedge  s_j\mydot\field{status} = \type{\%_ERR}~\big)\\
        &\textnormal{where}~ \type{\%}~ \textnormal{matches}~ \type{CHAL},~\type{CHALRESP} ~\textnormal{or} ~\type{RESP}\\[-0.4em]
 \end{split}
 \]
\end{llemma}%
\stepcounter{ctlem}%
\noindent
\cref{lem:security-low-M} states that if a (challenge, challenge-response, or response) packet is modified by a man-in-the-middle attack (adversarial behavior \cref{attack-chip}), {\system} should raise an error-specific alarm by visiting the corresponding $\type{ERR}$ state (after the $\type{ATTK}$ state) and finally terminate in the $\type{ABORT}$ state.

\begin{center}
	\begin{table}[t]
		\renewcommand{\arraystretch}{1.3}
		\caption{Proof efforts in Isabelle/HOL ($\sim$7,100 LoC).}\label{tab:proof}
		\begin{tabularx}{\linewidth}{ccccc}
			\toprule
			\multirow{2}{*}{\textbf{Level}} &  \multicolumn{2}{c}{\textbf{Specifications}} & \multicolumn{2}{c}{\textbf{Proofs}}\\
			\cmidrule(lr){2-3} \cmidrule(lr){4-5}
			& \#locale/definition & LoC & \#property/lemma & LoC\\
			\midrule
			$\spec_h$& 1 & $\sim$50 & 2 & $\sim$50\\
			$\spec_l$& 90 & $\sim$850 & 304 & $\sim$6,150\\
			\bottomrule
		\end{tabularx}
	\end{table}
\end{center}

\paragraph*{\textbf Mechanized proof in Isabelle/HOL}
In our specification $\spec_l$, an agent BSP/AP can extend a trace (i.e., a \emph{run} $\run$ of $\spec_l$) in any way permitted by the protocol. An adversarial behavior can also change the current state and extend a trace. 
Low-level properties on both functional correctness and security are formalized as trace properties and proved by induction on traces of $\spec_l$.
In particular, security properties \cref{lem:security-low-B,lem:security-low-M} both cover different possible traces of $\spec_l$ (corresponding to different attack scenarios). Therefore, we further decompose these lemmas into a set of \emph{auxiliary lemmas} to account for different traces and prove by induction on each trace.
Then, by chaining these auxiliary lemmas together in Isabelle/HOL, we obtain a fully mechanized proof that all possible traces of $\spec_l$ (i.e., all possible executions of {\system}) fulfill the requirements on functional correctness and security. \Cref{tab:proof} collects the statistics with regard to the proof efforts conducted in Isabelle/HOL, which amount roughly to 8 person-months.

\section{Implementation and Evaluation}\label{sec:evaluation}
This section presents {\csystem}, a proof-of-concept implementation of {\system} embedded in multiprocessor secure boot.
To build confidence in the consistency of the formal model and the implementation, we develop a test framework that extracts executable code from our Isabelle model $\spec_l$ and validate it against {\csystem}.
We then evaluate the security and performance of {\csystem} on ARM Fixed Virtual Platform (FVP) based on Fast Models 11.18 \cite{fvp1}. The tests are conducted on the Foundation Platform \cite{fvpmodel} with ARM Cortex-A72 CPUs, with CPU0 performing as BSP and CPU1 as AP.

\subsection{Proof-of-Concept Implementation}\label{subsec:impl}

We derive {\csystem} in C ($\sim$1,400 LoC) for the ARM64 bare-metal environment as an instance of the (low-level) formalization of {\system} in Isabelle/HOL. 
This instantiation is justified by a code-to-spec review in the same way as~\cite{zhao2017refinement}, establishing a near one-to-one correspondence between the formalization and the implementation, e.g.,
the correspondence between the events defined in $\spec_l$ (see \cref{subsec:low-spec}) and the C functions declared in {\csystem}. We build {\csystem} using GCC 9.4.0 cross-compiler on a server running 64-bit Ubuntu 20.04. {\csystem} is further embedded in bootwrapper v0.2 \cite{bootwrapper}, a bootloader for the ARMv8 architecture, to perform authentication at the initial boot stage. To use cryptographic primitives in the boot environment, {\csystem} adopts wolfSSL 5.3.0 \cite{wolfSSL}. Moreover, {\csystem} uses Newlib 4.3.0 \cite{newlib} as the C standard library. Functional correctness of {\csystem} is evaluated on the functional-accurate simulator FVP.


\paragraph*{\textbf {\csystem} bootloader}
We integrate {\csystem} at a later phase of bootwrapper such that necessary hardware initialization is finished before the execution of {\csystem}. We refer to the resulting bootloader as the \emph{{\csystem} bootloader}. Upon the launch of {\csystem}, both processors \emph{concurrently} validate the other processor's certificate until the initiation phase (cf.\ \cref{subsec:initiation}) is completed, i.e., when both processors have retrieved each other's public key and the validations are successful\footnote{The processor that finishes the validation earlier than the other halts and waits for the other processor.}. 
During the challenge-response phase, both processors \emph{sequentially} exchange encrypted packets (as the inter-processor communications cannot be parallelized) using asymmetric cryptography, then establish a shared session key (in the \emph{concurrent} mode) to secure the communication channel. Since FVP does not support inter-processor communication buses, the channel is mimicked by the main memory where both processors can read from and write to a pre-defined memory location. Moreover, {\csystem} uses Set Event (SEV) and Wait For Event (WFE) instructions for inter-processor synchronizations: after writing its packet to the channel, a processor executes a SEV instruction to wake the other, then immediately suspends itself by executing WFE.


{\csystem} uses ECC for its asymmetric cryptographic operations due to its efficiency and security over RSA. It employs elliptic-curve digital signature algorithm (ECDSA) for certificate generation and validation, elliptic-curve Diffie-Hellman (ECDH) for secure key exchange, and ephemeral ECDH (ECDHE) for session key generation to ensure forward secrecy.
An ephemeral ECC key pair is generated using the P-256 curve, producing a 256-bit private key and its corresponding public key. The shared session key, derived via ECDHE between BSP and AP, is also 256 bits. In the challenge-response phase, each party computes a shared secret using ECDH with its private key and the other party’s public key.
This shared secret is processed with HKDF-SHA256 to derive cryptographic keys. AES-128-CBC is used for data encryption, and HMAC-SHA256 ensures message authentication. These operations, including HKDF-SHA256 and AES-128-CBC, are internally invoked by the wolfSSL library during ECDH execution.
Private keys, certificates, and hash values are linked to the {\csystem} bootloader during compilation and loaded into memory alongside it.

\paragraph*{\textbf Library compatibility}
We tune compile-time configurations of the invoked libraries for compatibility with the bare-metal environment: bootwrapper, Newlib, and wolfSSL are compiled with the flag $\flag{-mstrict-align}$ to disable unaligned accesses;
%
For Newlib, we adapt bootwrapper's linker script to make memory-management functions such as $\func{malloc}$ and $\func{free}$ work correctly, e.g., to avoid heap-memory conflicts; 
%
For wolfSSL, we define compile-time macros to fit for the bare-metal environment, which mainly include
\begin{itemize}[leftmargin=16pt,labelwidth=8pt,labelsep=6pt]
    \item $\flag{NO\_FILESYSTEM}$ to disable loading of keys and certificates in the system's file system,
    \item $\flag{WOLFSSL\_USER\_IO}$ to remove automatic setting of default I/O functions, and
    \item $\flag{NO\_WRITEV}$ to disable $\func{writev}$ semantics simulation.
\end{itemize}
We further add macros to enable all ECC-related operations. The libraries are statically linked to {\csystem} bootloader as the bare-metal environment does not support dynamic load.

\subsection{Validation Framework}\label{subsec:valid}
Here, we present the process to validate the consistency between the C implementation and Isabelle specification. 
A test framework is developed to extract executable OCaml code from our Isabelle model $\spec_l$ and validate it against {\csystem}, covering both normal protocol execution and attack scenarios.

\begin{figure}
    \centering
    \includegraphics[width=\linewidth]{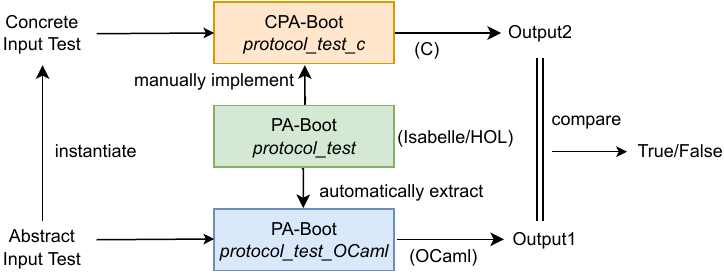}
    \caption{Validation framework}
    \label{fig:valid}
\end{figure}
\paragraph*{\textbf Consistency validation}
The validation is performed using an executable version generated in OCaml via Isabelle/HOL's built-in extraction mechanism.
Based on the extracted OCaml code, as illustrated in \cref{fig:valid}, our validation framework operates as follows.
Our focus is on test cases that simulate normal protocol execution as well as the actions of a skilled attacker carrying out real attack effects, as described in \cref{subsec:sec-eval}.
Given an abstract input test case, the extracted OCaml function $\field{protocol\_test\_OCaml}$ executes the test case and produces an output.
Simultaneously, we instantiate the input test case as corresponding concrete ones, and {\csystem} produces an output.
If the two results match, the test is valid, demonstrating consistency between the two implementations. If the results differ, the framework identifies an inconsistency between the low-level C code and OCaml code (and, by extension, the high-level Isabelle/HOL model).

The OCaml implementation extracted from the formal model acts as a higher-level abstraction of the C implementation.
For instance, private keys are abstractly represented as an enumerated type with cases of \type{ROOT_PRIVK}, \type{M_PRIVK}, \type{S_PRIVK} and \type{BAD_PRIVK} in Isabelle/HOL and thereby in OCaml. The OCaml side uses \type{BAD_PRIVK} to represent a scenario involving a tampered AP. In contrast, the C side simulates an AP-replacement attack by statically linking a modified private key file of CPU1.

In the OCaml implementation, the output result indicates whether the protocol reaches the terminal state with an \type{END} status or \type{ABORT} status. Similarly, in {\csystem}, the result reflects whether the protocol executes successfully and progresses to the next boot stage, or aborts with a detected attack.
The results demonstrate the consistency of the outputs for matching inputs across various scenarios particularly those covered in \cref{tab:attack}, thereby increasing confidence in the accuracy of our C implementation.

\subsection{Security Evaluation}\label{subsec:sec-eval}
\begin{table}
    \setlength\tabcolsep{3pt}
    \centering
    \caption{ 
    Instances of adversarial behaviors simulated via FVP. Acronyms: $\acronym{APR}$ for AP replacement, $\acronym{RCT}$ for root-certificate tampering, $\acronym{APCT}$ for AP-certificate tampering, $\acronym{CPM}$ for challenge-packet manipulation, $\acronym{CRPM}$ for chal.-resp.-packet manipulation, and $\acronym{RPM}$ for response-packet manipulation.}
    \label{tab:attack}
    \renewcommand{\baselinestretch}{1.0} 
    \begin{tabular}{lccccccccc}
        \toprule
        \textbf{Category} & &\multicolumn{3}{c}{\cref{attack-AP}} & & \multicolumn{3}{c}{\cref{attack-chip}} & \\
        \cmidrule(lr){3-5} \cmidrule(lr){7-9}
        \textbf{Instance} & & $\acronym{APR}$ & $\acronym{RCT}$ & $\acronym{APCT}$ && $\acronym{CPM}$ & $\acronym{CPRM}$ & $\acronym{RPM}$ & \\
        \textbf{Detected} & & \cmark & \cmark & \cmark && \cmark  & \cmark & \cmark\\
        \bottomrule
    \end{tabular}
\end{table}

We perform empirical security evaluations of {\csystem} to detect aforementioned adversarial behaviors simulated in FVP covering \cref{attack-AP} and \cref{attack-chip}. In particular, we introduce an extra processor CPU2 mimicking the interposer chip to launch man-in-the-middle attacks. As summarized in \cref{tab:attack}, {\csystem} succeeds in detecting all different instances of these adversarial behaviors where the BSP returns error-specific alarms to abort the boot process. We show below how the adversarial behaviors are implemented and how {\csystem} detects them. 

\paragraph*{\textbf AP replacement}
We modify the private key $\privkap$ of CPU1 to simulate the AP-replacement attack.
Such attack does not trigger alarms in certificate validation and {\csystem} enters the challenge-response phase. In this phase, CPU1 attempts to forge packets to pass the authentication. However, since the modified private key of CPU1 does not match the stored certificate $\cert{AP}$ that has been validated in initiation phase, CPU0 cannot decrypt the challenge packet and hence raises an AP-replacement error and aborts the boot process. 

\paragraph*{\textbf Man-in-the-middle attacks}
CPU2 attempts to eavesdrop on or tamper with the (challenge, challenge-response or response) packets transmitted over the inter-processor communication channel (mimicked by memory) during protocol execution. Particularly, for the latter two types of packets, CPU2 acts as a skilled attacker who attempts to replace $\epubk{BSP}$ in the challenge-response packet and $\epubk{AP}$ in the response packet aiming to establish a shared session key for future communications. However, as the attacker does not know the private keys of CPU0 and CPU1, he/she cannot manipulate the encrypted hash values of ephemeral public keys. As a consequence, {\csystem} observes unmatched hash values and thus raises an alarm and aborts the boot process.

\paragraph*{\textbf Tampering with certificates}
We modify certificates linked to {\csystem} bootloader to simulate certificate manipulations. Specifically, the attacker tries to modify the root certificate or the AP certificate to bypass the processor authentication in the challenge-response phase. However, since the attacker cannot modify the hash value of the root certificate stored in the bootROM, CPU0 detects the root-certificate manipulation using the hash and terminates the boot process. For the AP-certificate manipulation, CPU0 uses the validated root certificate to verify the manipulated AP certificate (signed by $\certroot$) which leads to abortion as well.

\begin{table}[t]
		\renewcommand\arraystretch{1.3}\footnotesize
		\caption{\#instructions consumed by the main boot stages, i.e., bootwrapper (b.w.p.), {\csystem}, and Linux kernel boot.}
		\centering
		\label{tab:stage-cycle}
		\begin{tabular}{clllc}
			\toprule
			\textbf{Processor(s)} &  \!\!\!\!\!\:\makecell[c]{\textbf{b.w.p.}\\($\alpha$)} & \!\!\!\:\makecell[c]{\textbf{\csystem}\\($\beta$)} & \!\!\!\!\:\makecell[c]{\textbf{Kernel boot}\\($\gamma$)} & \!\!\!\!\makecell[c]{\textbf{Overhead}\\($\nicefrac{\beta}{(\alpha+\gamma)}$)}\\ 
			\midrule
			CPU0 & \!\!\;4,939 & \;52,091,762 & \!\;1,683,639,749 & \!\!\!\!3.09\% \\ 
			CPU1 & \!\!\;\phantom{4,}650 & \;52,181,244 & \!\;1,759,591,720 & \!\!\!\!2.97\% \\ 
			CPU0 $\parallel$ CPU1 & \!\!\;4,939$^\star$ & \;87,570,653$^\diamond$  & \!\;1,759,591,720$^\star$ & \!\!\!\!4.98\% \\ 
			\bottomrule\\[-1em]
		\end{tabular}
	{\scriptsize $\alpha$, $\beta$, or $\gamma$ denotes the number of instructions executed at each stage, respectively, e.g., $\alpha_{\text{CPU0}\,\parallel\,{\text{CPU1}}}$ for bootwrapper executed instructions of the dual-processor system. \\
    $^\star \alpha_{\text{CPU0}\,\parallel\,{\text{CPU1}}} = \max\{\alpha_{\text{CPU0}}, \alpha_{\text{CPU1}}\}$ because bootwrapper is assumed to be fully parallelized. The same applies to the boot stage of the Linux kernel.\\
	$^\diamond \max\{\beta_{\text{CPU0}}, \beta_{\text{CPU1}}\} < \beta_{\text{CPU0}\,\parallel\,{\text{CPU1}}} < \beta_{\text{CPU0}} + \beta_{\text{CPU1}}$ since {\csystem} is partially parallelized, i.e., communications in the chal.-resp.\ phase must be in sequence.
	}
 \vspace{1em}
\end{table}

\begin{figure}[t]
	\centering
	\hspace*{-3mm}
	\begin{tikzpicture}
		\pgfplotsset{every tick label/.append style={font=\tiny},
			every x tick scale label/.append style={yshift=1.4em, xshift=1.54em},
			every axis./.append style={thick}
		}
		\begin{axis}[
			xbar stacked,
			legend style={
				nodes={scale=0.75, transform shape},
				legend columns= 2,
				at={(.77,.95)},
				anchor=north,
				draw=none
			},
			legend cell align={left}, 
			ytick=data,
			tick label style={font=\footnotesize},
			xtick style={
				/pgfplots/major tick length=3pt,
			},
			ytick style={
				/pgfplots/major tick length=0pt,
			},
			legend style={font=\footnotesize},
			label style={font=\footnotesize},
			width=.5\textwidth,
			bar width=6mm,
			xlabel={\textbf{\#instructions}},
			yticklabels={CV0, CV1, CR0, CR1},
			ytick=data,
			area legend,
			y=8mm,
			ymin=0,
			xmin=0,
			enlarge y limits={abs=0.75},
			]
			\addplot[fill=white,draw=figred, postaction={pattern=crosshatch,pattern color=figred}] coordinates
			{(10393204,3) (10659064,2) (0,1) (0,0)};
			\addplot[fill=white,draw=CertsValid,postaction={pattern=north west lines,pattern color=CertsValid}] coordinates
			{(0,3) (0,2) (13656907,1) (6870226,0)};
			\addplot[draw=figyellow2, fill=figyellow] coordinates
			{(0,3) (0,2) (6794376,1) (13820184,0)};
			\addplot[draw=ForestGreen, fill=figgreen!60] coordinates
			{(0,3) (0,2) (19944663,1) (19996249,0)};
			\addplot[draw=white,fill=none] coordinates{(0,3) (0,2) (0,1) (0,0)};
			\addplot[draw=figgrey2,fill=figgrey] coordinates{(271534,3) (272002,2) (1003116,1) (919465,0)};
			\legend{ECCValid, ECCEnc, ECCDec, ECDHE, \phantom{ghost},Others}
		\end{axis}
	\end{tikzpicture}
	\caption{\#instructions by different operations in {\csystem}. CV0 denotes certificate validation in the initiation phase performed by CPU0; CR0 denotes the challenge-response phase on CPU0. The notations for CPU1 are analogous. 
		ECCValid: ECC-based certificate verification;
		ECCEnc/ECCDec: ECC-based public-key encryption/decryption via the ECDH algorithm;
		ECDHE: ECC-based key exchange, including the generation of ephemeral key pairs and $\sessk$;
		Others: memory read/write, SHA-256 processing, etc.
	}
	\label{fig:step-cycle}
\end{figure}
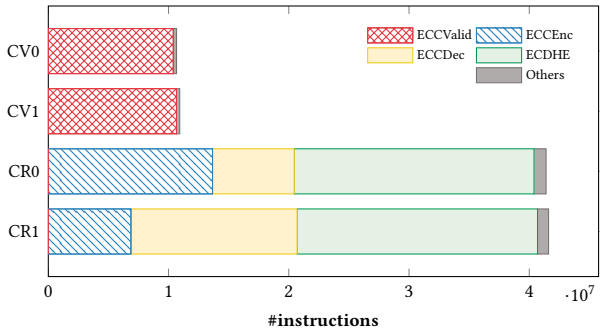

\subsection{Performance Evaluation}\label{subsec:perf-eval}

we report the performance of {\csystem} within a complete boot process from bootloader to shell login on FVP. We measure the number of executed instructions rather than CPU cycles of the boot process, as FVP is not cycle-accurate and each instruction takes equally one cycle to execute \cite{fvp-cycle}. 
To capture this, we enable performance monitor unit (PMU \cite{pmu}) monitoring by setting a PMU control register at key points in the boot stages, recording the total number of instructions executed on FVP. 
%
%
To assess the impact of {\csystem} on boot performance, we additionally measure the boot process of a Linux system, namely, the Gentoo Linux distribution (stage archive 3, with systemd as init system, kernel version 5.16.0) \cite{gentoo}.
%
%
Moreover, we disable Linux kernel PMU support to avoid kernel's modification on the PMU counters.

\cref{tab:stage-cycle} quantifies the experimental performance 
of the entire boot process averaged over three boot trials, consisting of three main boot stages, i.e., bootwrapper, {\csystem}, and Linux kernel boot. 
The integration of {\csystem} adds a 4.98\% overhead to the multiprocessor secure boot, roughly 3\% per CPU, which is relatively small. 
Moreover, we report the performance of different operations conducted in {\csystem}. As depicted in \cref{fig:step-cycle}, the main overhead of {\csystem} stems from asymmetric cryptography used for certificate validation, packet encryption and decryption, and key exchange.

\section{Discussion and Future Work}\label{sec:discussion}
In this section, we briefly discuss some issues pertinent to our approach and several interesting future directions.

\paragraph*{\textbf{Hardware Adaptation}}
We implemented our protocol on FVP to demonstrate its security and usability. Here, we discuss how to deploy it on real hardware.
Based on the analysis of required components (highlighted in yellow in \cref{fig:mp-arch2}), we identify necessary hardware modifications during manufacturing. The primary requirement is customizing the CPU during manufacturing, specifically by embedding the {\system} protocol code into the immutable bootROM. This is feasible since OEMs typically collaborate with silicon vendors for such modifications. Additionally, the root certificate hash must be stored in the bootROM for processor certificate validation. 
The OEM must modify the NVM to store certificates, which is easily reprogrammable.
The generated session key does not require hardware modifications and can be stored in a CPU register to secure communication during the subsequent boot process.
As noted in \cref{subsec:threat}, our approach is particularly useful when the OEM is located far from the end-user, such as overseas, and lacks full control over the supply chain. This is essential for devices performing critical tasks, such as data center servers that support enterprise workloads like cloud services and big data processing.

\paragraph*{\textbf{Protocol Extensions}}
We envision three key extensions:
\begin{enumerate*}[label=(\roman*)]
\item {\system} applies to scenarios with multiple APs, allowing the BSP to authenticate them sequentially. This aligns with industrial practices, where systems in critical or embedded environments often perform sequential startups where BSP wakes up APs in sequence, to ensure control and simplify debugging. Specifically, {\system} integrates into the boot process using a dedicated IPI channel for inter-processor communication. After hardware initialization, APs enter a suspended state, awaiting a startup signal from the BSP. The BSP activates each AP individually by setting its APIC ID in the APIC's ICR register and then authenticates it.
In contrast, simultaneous startups involve the BSP broadcasting signals to initialize all APs concurrently, resulting in parallel authentication sessions. Addressing this scenario is part of our future work.
\item %
Another direction to extend our protocol is to support a dual-return approach. Users can choose a permissive mode, which allows the boot process to continue despite certain failures, and a strict mode, which aborts the process on any error. 
In scenarios with multiple APs, the permissive mode can disable APs that fail verification, and allowing the boot process to continue.
Since {\system} executes before loading the bootloader or kernel image--when the system is vulnerable to attacks that could modify these components by a malicious AP--disabling the APs at the end of {\system} prevents such attacks.
%
\item Another potential extension is to incorporate bootROM integrity into the AP’s key derivation process, protecting against very powerful attackers who can tamper with the bootROM. For example, the protocol could employ Physically Unclonable Functions (PUFs) to dynamically generate keys at runtime.
The PUF response could be combined with the bootROM hash (e.g., using XOR or a key derivation function) to form the key.
This ensures that any modification to the bootROM alters the AP’s derived private key, allowing our protocol to detect bootROM compromise through a mismatch with the stored certificate $\cert{AP}$.

\end{enumerate*}


\paragraph*{\textbf C code generation}
Our proof-of-concept prototype is implemented in C to leverage its efficiency.
While Isabelle/HOL's can generate executable code (e.g., in OCaml), this code is not optimal in terms of performance.  
A promising direction for future work is to derive verified C code. 
One approach is to extend Isabelle's toolchain to directly synthesize C code, as has been observed in~\cite{DBLP:journals/corr/abs-1812-03318}. 
To achieve this, we plan to refine our low-level specification in Isabelle/HOL using the Simpl language~\cite{DBLP:phd/de/Schirmer2006} (with C-like syntax) and develop a verified compiler from Simpl to C within Isabelle/HOL.
Another option is to use the Igloo framework \cite{sprenger2020igloo} to model our protocol and extend it to support our security properties. 
This would allow us to translate Isabelle's formal specifications into program assertions, which can then be verified with existing tools to ensure the C implementation meets the specified properties.
Alternatively, we could rewrite our protocol in Low* and use the KaRaMeL compiler \cite{karamel} to generate certified C code.

\section{Related Work}\label{sec:relatedWork}

This section reviews research closely related to our approach, focusing on formal methods in bootstrap authentication and protocol verification.
\paragraph*{\textbf{Secure Boot/Authenticated Load Verification}}
Formal methods have witnessed a spectrum of applications in enhancing confidence in boot code security \cite{straznickas2020towards,cook2020model,huang2018formal}.
More pertinently to our work, Muduli \etal. \cite{DBLP:conf/fmcad/MuduliSR19} verified the end-to-end security of authenticated firmware loaders using model-checking techniques \cite{DBLP:books/daglib/0020348}.
The targeted scenarios and security properties are different from ours. Muduli \etal. focuses on vulnerabilities such as protocol-state hijacking, time-of-check to time-of-use (TOCTOU), and confused deputy attacks \emph{during} firmware loading, whereas our protocol ensures processor authenticity and inter-processor communication confidentiality \emph{before} firmware loading. Similar differences apply to \cite{6855571}, which models firmware loading flows in Promela and verifies the absence of TOCTOU attacks using the Spin model checker \cite{DBLP:journals/tse/Holzmann97}.
Cremers \etal. \cite{cremers2023formal} formalized the SPDM 1.2 protocol standard in the Tamarin prover \cite{DBLP:conf/csfw/SchmidtMCB12}. SPDM aims for device attestation, authentication and secure communication and can be used during the boot process. However, due to its size and complexity, the formalization is not fully verified. Although they split the protocol into four separate models and verified them separately, they do not analyze cross-protocol attacks or verify security properties on the complete model.
%
%
In summary, no existing work addresses defending against the hardware supply-chain attack surface, as identified in \cref{subsec:threat}.

\paragraph*{\textbf Protocol verification}
Formal methods have been widely applied to ensure security of real-world protocols, like TLS 1.3 \cite{DBLP:conf/sp/CremersHSM16}, messaging protocols \cite{DBLP:conf/eurosp/KobeissiBB17}, and entity authentication protocols \cite{DBLP:conf/ccs/BasinDHRSS18,DBLP:conf/ccs/SprengerB10}.
There are dedicated tools for protocol verification, such as Tamarin \cite{DBLP:conf/csfw/SchmidtMCB12} and ProVerif \cite{DBLP:conf/csfw/Blanchet01}.
However, protocols verified in these frameworks are abstract models that may not guarantee the same properties in actual implementations \cite{DBLP:conf/sp/ArquintWLSSWBM23}. 
Works using F* \cite{swamy2011secure}, Coq, and Isabelle/HOL seek to address this limitation.
Protocols in F* can be compiled into OCaml or F\# code, while protocols in Low* \cite{protzenko2017verified}, a low-level subset of F*, can be translated into C via the KaRaMeL compiler. 
Low* supports the development of high-assurance cryptographic libraries, such as HACL* \cite{zinzindohoue2017hacl}, and implementations of protocols like TLS 1.3 \cite{delignat2017implementing} and protocols for measured boot \cite{protzenko2019formally} or bootloaders \cite{yuan2021verified}.
The DY* symbolic protocol framework \cite{bhargavan2021text}, written in F*, combines the precision of dependent types from F* with the automation of symbolic execution in dedicated provers like Tamarin, offering efficient and expressive verification.

%
The following discusses the most closely related works that formalize protocols using Isabelle/HOL and probably derive verified implementation.
%

Paulson's seminal work\cite{DBLP:journals/jcs/Paulson98, DBLP:journals/tissec/Paulson99} uses an inductive approach to model network protocols and prove their security properties.
However, it does not verify our security properties \cref{lem:security-low-B} and \cref{lem:security-low-M},  
which ensures that once an adversarial behavior is performed during protocol execution, the protocol should detect it at some later state along the trace, and ultimately terminate in an error state and trigger an alarm.
These properties are crucial for identifying attack vectors and preventing attackers from gaining control of the system.

Igloo \cite{sprenger2020igloo} also uses Isabelle/HOL with multiple specification layers and guides the correctness proof of implementation code but follows a different approach. It establishes a compositional refinement framework using forward simulation to prove refinement between layers.
The soundness theorem establishes that refinement implies trace inclusion.
Igloo shows that the abstract protocol model satisfies a trace property, while the lower layer, though introducing more events, refines the higher layer and preserves the trace property.
In contrast, our work leverages Isabelle’s built-in locale module for abstraction-refinement reasoning. 
Unlike Igloo, which is based on Paulson’s approach, our work verifies security properties \cref{lem:security-low-B} and \cref{lem:security-low-M}.
While both link high-level models to low-level implementations, Igloo manually converts formal requirements defined in Isabelle into program specifications annotated in programs and uses tools like VeriFast for verification. Our approach prioritizes automation, ensuring consistency between formal specification and implementation through extracted executable code and testing.

\section{Conclusion}\label{sec:conclusion}

We have identified a new, prevalent hardware supply-chain attack surface that can bypass multiprocessor secure boot due to the absence of processor-authentication mechanisms. To defend against these attacks targeting \enquote{assumed-safe} components (e.g., processors and inter-processor communication channels), we presented {\system}, the first formally verified processor-authentication protocol for multiprocessor secure bootstrap. We showed -- using a machine-checked mathematical proof in Isabelle/HOL -- that {\system} is functionally correct and is guaranteed to detect multiple adversarial behaviors, e.g., man-in-the-middle attacks and processor replacements. Experiments on ARM FVP suggested that our proof-of-concept implementation {\csystem} can effectively identify boot-process attacks with a minor overhead and thereby improve the bootstrap security of multiprocessor systems.

Future work includes expanding the functionality of {\system} and generating verified C code (see \cref{sec:discussion}).

\bibliographystyle{IEEEtran}
\bibliography{main}

%
%
%





\end{document}